\documentclass[structabstract]{aa}  
	\usepackage{natbib}                                   		
	\bibpunct{(}{)}{;}{a}{}{,}	
	\usepackage{graphicx}
	\usepackage{txfonts}
\begin{document}
	\authorrunning{Gonz\'alez \and Reisenegger}
	\titlerunning{Internal heating of old neutron stars}
	\title{Internal heating of old neutron stars: contrasting different mechanisms}
	\author{Denis Gonzalez \and Andreas Reisenegger}
	\institute{Departamento de Astronom\'\i a y Astrof\'\i sica, Pontificia Universidad
	Cat\'olica de Chile, Casilla 306, Santiago 22, Chile.\\
	\email{[dhgonzal;areisene]@astro.puc.cl}
	}
	\date{Received ; accepted }
	\abstract
%	context
	{The standard cooling models of neutron stars predict temperatures of $T<10^{4}$~K for ages $t>10^{7}$~yr. 
	However, the likely thermal emission detected from the  millisecond pulsar J0437-4715, of spin-down age 
	$t_s \sim 7\times10^9$~yr, implies a temperature  $T\sim 10^5$~K. Thus, a heating mechanism needs to be 
	added to the cooling models in order to obtain agreement between theory and observation.}
%	aims
	{Several internal heating mechanisms could be operating in neutron stars, such as magnetic field decay, dark matter accretion, 
	crust cracking, superfluid vortex creep, and non-equilibrium reactions  (``rotochemical heating''). We study these mechanisms 
	to establish which could be the dominant source of thermal emission from old pulsars.}
%	methods
	{We show by simple estimates that magnetic field decay, dark matter accretion, and crust cracking are unlikely	to have a 
	significant heating effect on old neutron stars. The thermal evolution for the other mechanisms is computed with the code of 
	Fern\'andez and Reisenegger. Given the dependence  of the heating mechanisms on the spin-down parameters, we study the thermal 
	evolution for two types of pulsars: young, slowly rotating  ``classical'' pulsars  and old, fast rotating millisecond pulsars.}
%	results
	{We find that magnetic field decay, dark matter accretion, and crust cracking do not produce any detectable heating of old pulsars. 
	Rotochemical heating and vortex creep can be important both for classical pulsars and millisecond pulsars. More restrictive
	upper limits on the surface temperatures of classical pulsars could rule out  vortex creep as the main source of 
	thermal emission. Rotochemical heating in classical pulsars is driven by the chemical imbalance built up
	during their early spin-down, and is therefore strongly sensitive to their initial rotation period.}
	{}
	\maketitle

	%%%%%%%%%%%%%%%%%%%%%%%%%%%%%%%%%%%%%%%%%%%%%%%%%%%%%%%%%%%%%%%%%%%%%%%%%%%%%%%%%%%%%%%%%%%%

	\section{Introduction}
	\label{sec:intro}	
	Neutron stars (NSs) are compact objects composed of a  liquid core enveloped by a solid crust. The core is expected 
	to contain superfluid neutrons and superconducting protons, while the crust contains heavy atomic nuclei arranged 
	in a crystal lattice,  coexisting with superfluid neutrons in its inner part. The high density of the NS core, up to 
	$\sim(3-9)\rho_0$, where $\rho_0$ is  the saturation nuclear matter density, cannot be reproduced in terrestrial laboratories. 
	This turns the NSs into natural laboratories. The study of their thermal evolution, confronting theory and observation, 
	provides a useful test for the understanding of the properties of matter at supernuclear density.

	For all standard cooling models \citep{yak04}, neutron stars cool down to surface temperatures $T_s < 10^4$~K 
	within less than $10^7$~yr. Nevertheless, the observation of ultraviolet thermal emission from millisecond pulsar
	J0437-4715 \citep{kar04}, whose spin-down age, corrected to the latest distance of $157$~pc \citep{del08}, is 
	$\tau_{sd}\sim 7\times 10^9$~yr \citep{van01}, shows a surface temperature of about $\sim 10^5$~K for this 
	pulsar. Hence, a heating mechanism needs be to added  to the standard cooling models  to obtain agreement between 
	theory and observation.

	There are several heating mechanisms that can be present during the late stages of the thermal evolution. These include the 
	frictional motion of superfluid neutron vortices \citep{alp84,shi89,lar99}, rotochemical heating 
	\citep{reis95,reis97,fer05,petro09}, magnetic field decay \citep{gol92,thom96,pon07}, and crust cracking \citep{baym71,chen92}. 
	Other mechanisms, based on more speculative hypotheses, such as a time variation of the gravitational constant \citep{jof06}, the
	decay of exotic particles \citep{han02}, or the accretion of dark matter particles \citep{dela10,kouv10}, could in principle also 
	heat old neutron stars. In \citet{sch99} and \citet{lar99}, several of the internal heating mechanisms cited are studied and
	confronted with observational data of neutron stars (surface temperature upper limits). However, the oversimplified description 
	of some heating mechanisms \citep[rotochemical heating in][]{sch99}  and the until then non-detection of thermal 
	emission from neutron stars older than $\sim10^6$ yr, made it impossible to obtain reliable conclusions.

	The goal of this work is  to provide a comparative analysis of the thermal evolution including different heating  mechanisms. 
	In order to do this, we discard some of them as not strong enough (magnetic field decay, dark matter accretion, and
	crust cracking), and we present a more detailed study of the most promising ones: vortex creep and rotochemical heating. We 
	confront these mechanisms with  the thermal emission detected in the millisecond pulsar J0437-4715 and the best available  
	upper limits  on the temperature of  six other old pulsars. Owing to the dependence of heating mechanisms on spin-down 
	parameters, which leads to different temperatures for different pulsars, we separately study the thermal evolution for two 
	types of pulsars: young, strongly magnetized, and slowly rotating ``classical'' pulsars, and old, weakly magnetic, and fast 
	rotating millisecond pulsars (MSPs). 

	The paper is organized as follows. In Sect.~\ref{sec:mech} we describe the different heating 
	processes and assess their importance. In Sect.~\ref{sec:eff} we show the effects of vortex creep and rotochemical heating in 
	classical and millisecond pulsars. In Sect.~\ref{sec:obs} the predictions are confronted with observed data. A summary of our 
	main conclusions is given in Sect.~\ref{sec:conc}.

\section{Mechanisms}\label{sec:mech}

	The cooling  of a  NS is caused by neutrino emission from the interior and by thermal photon emission from the surface of the 
	star. The neutrino emission is determined by the state of matter in the core, which depends on the stellar mass and the 
	properties of the matter at nuclear density. The evolution of the internal temperature of an NS is given by the thermal balance 
	equation
	\begin{equation}
	\label{eq:ther_bal}
	\dot T= \frac{1}{C} ( L_H - L_{\gamma} - L_{\nu}),
	\end{equation}
	where $C$ is the total heat capacity of the star, $L_{\gamma}$ is the photon luminosity, $L_{\nu}$ is the neutrino luminosity, 
	and $L_H$ is the power generated by internal heating mechanisms, of which we consider the following.

\subsection{Magnetic field decay}

	\citet{gol92}  studied the processes that promote the dissipation of magnetic energy in  NSs. Later, \citet{thom96}  studied 
	the resulting emission (X-rays, neutrinos, Alfv\'en waves)  from very strongly magnetized NSs (``magnetars''), and \citet{pon07} showed observational evidence that suggests 
	heating by this mechanism in  different classes of relatively  young, strongly magnetic NSs. Because the physical processes  involved in the decay
	of  NS  magnetic 
	fields are still uncertain, we  only make an order-of-magnitude estimate of the magnetic field $B$ required to produce a detectable 
	surface temperature in old NSs.
	
	At ages  $t>10^6$~yr, cooling  is dominated by photon emission. The luminosity caused by the decay of the magnetic
	field in a NS with radius $R$ and  magnetic energy   
	$E_B\sim (4\pi R^3/3)\langle B^2\rangle/8\pi$  in a time scale  $t$, is 
	\begin{equation}
	L= 4\pi R^2 \sigma T_s^4 \sim \frac{E_B}{t}\sim \frac{4\pi R^3}{3}\frac{\langle B^2\rangle}{8\pi}\frac{1}{t}.
	\nonumber
	\end{equation}
	In this way, the magnetic field required to account for an NS with a surface temperature $T_s\sim 10^5$~K (i.e. the 
	detected temperature in  J0437-4715) and radius $R=10$~km is 
	\begin{equation}
	B_{rms}\equiv\sqrt{\langle B^2 \rangle}\sim  \left(\frac{24\pi\sigma T_s^4 t}{R}\right)^{1/2}=10^{13} \sqrt{t_{7}}\textrm{ G},
	\end{equation}
	where $t_7$ is the age of the NS in units of $10^7$~yr. An old classical pulsar of $\sim 10^7$~yr and 
	an MSP of $\sim 10^9$~yr require, respectively, magnetic fields of $\sim 10^{13}$~G and 
	$\sim 10^{14}$~G in order to obtain a detectable temperature\footnote{These values might be reduced if a substantial fraction
	of the protons is the star are superconducting \citep{eas77}.}. 
	Thus, the magnetic fields inferred from the spin-down  
	in classical pulsars, $\sim 10^{11}$~G, and MSPs,  $\sim 10^8$~G, are  much too low to produce  detectable heating.

\subsection{Dark matter accretion}
	Recently, \citet{kouv10} and \citet{dela10} studied the effects of dark matter (DM)  accretion and annihilation in compact 
	objects. One of these is the potential increase of the surface temperature of old NSs in sufficiently dense  environments (e.g., 
	the Galactic center and globular clusters). Because we are interested in the pulsars of  the solar neighborhood, where  
	the DM density is relatively low, we only make an order-of-magnitude estimate of the maximum  possible thermal 
	emission caused by this mechanism. The maximum possible accretion rate onto a NS is given by
	\begin{equation}
	\dot M \approx \rho_{DM} v_{\infty} \pi b_{\infty}^2,
	\end{equation}
	where $\rho_{DM}$ is the DM density, $v_{\infty}$ is the relative velocity between DM particles and the NS,
	and $b_{\infty}$ is the maximum impact parameter for a DM particle to hit NS. This is valid as long as the DM-baryon (or 
	DM-lepton) cross section $\langle\sigma\rangle \gtrsim \pi R^2/N_{bar} \sim 10^{-45}~\mathrm{cm}^2$, so any DM particle entering
	the NS stays inside. If $\langle \sigma \rangle$ is smaller than this, $\dot M$ is reduced correspondingly. From the Milky Way 
	(MW) rotation curve, the DM density in the solar neighborhood $\rho_{DM}\sim 10^{-2}~M_{\odot}~\mathrm{pc}^{-3}$. From Newtonian
	energy and angular momentum conservation, the  maximum impact parameter is $b_{\infty}\approx R v_{esc}/v_{\infty}$, in the 
	limit $v_{esc}\gg v_{\infty}$. Considering the escape velocity $v_{esc}\sim 2c/3$ and the typical velocities for NSs moving 
	through the MW,  $v_{\infty}\ga 2\times 10^7\mathrm{cm~s}^{-1}$, we can obtain an upper limit on the DM accretion rate, 
	$\dot M \la 9\times 10^{-25}~M_{\odot}~\mathrm{yr}^{-1}$. Hence, the maximum luminosity is 
	$\dot M GM/R\approx  10^{22}~\mathrm{erg~s}^{-1}$  for stable DM , and 
	$\dot M c^2\approx 5\times 10^{22}~\mathrm{erg~s}^{-1}$ for decaying DM. Considering that the cooling is dominated by photon emission,
	these correspond to the surface temperatures $\sim 2\times 10^3$~K and $\sim 3\times 10^3$~K for stable and unstable matter, 
	respectively. In order to get a detectable $T_s\sim 10^5$~K, a $10^6$ times higher DM density is required (as perhaps found 
	very near the Galactic center). Thus DM accretion does not have a significant effect in the solar neighborhood.

\subsection{Crust cracking}

	This mechanism considers a rotating NS whose  crust solidifies with an ellipsoidal form. The spin-down causes a gradual change 
	to a more spherical shape. In this process, the stress in the crust of the NS is 
	increased. However, the rigidity of the crust causes it to remain more oblate than it would be if it had no resistance to shear. 
	When the crust reaches a critical deformation, it breaks, part of the accumulated strain energy is released, and the excess 
	oblateness, due to the crust rigidity, is reduced. The mean stress $\sigma$ in the crust caused by the spin-down is given by 
	$\sigma = \mu (\epsilon_0 - \epsilon)$ \citep{baym71}, with $\epsilon=(I-I_0)/I_0$, where $\mu$ is the mean shear modulus of 
	the crust, $I$ is the moment of inertia of the star, $I_0$ is its non-rotating value, and $\epsilon_0$ is  a value of 
	$\epsilon$  at which the crust is stress-free.
	As the stellar rotation slows down, $\epsilon$ is correspondingly  decreased, increasing the accumulated stress. Eventually,
	the crust cracks, $\epsilon_0$ is suddenly decreased, causing a discrete change $\Delta(\epsilon_0-\epsilon)$.
	 The amount of strain energy released by this crack is \citep{baym71}
	\begin{equation}
	\Delta E_{str} = -2 B (\epsilon_0 - \epsilon)\cdot \Delta( \epsilon_0 - \epsilon),
	\end{equation}
	and the time between successive  quakes is
        \begin{equation}
        \Delta t\simeq 2A \frac{ \Delta(\epsilon_0-\epsilon)}{\Omega\dot\Omega}\left(\frac{\partial I}{\partial \epsilon}\right)^{-1},
        \end{equation}
	where $\Omega$ is the angular velocity of the star,
	 $B\sim  \mu V_c/42 $ (with $V_c$ the crust volume), and $A$ quantifies the increase in gravitational energy  due to the deformation of the shape 
	(for more detail, see \citealt{cut03}; \citealt{zdu08}). 
	If $\epsilon_0-\epsilon\approx\theta_c$, the critical breaking strain angle, and the time between quakes 
	is small compared to the timescale of thermal evolution,  the  time-averaged energy dissipation  rate by this process is
	\begin{equation}
	L_{cc}=bI\theta_c\Omega |\dot \Omega|,
	\end{equation}
	where $b= B/A$. This expression, based on the formalism of \citet{baym71}, differs 
	from that of \citet{chen92}, but  agrees with the correction made by \citet{sch99} for the particular case of a neutron star 
	modeled as a Maclaurin spheroid.

	\citet{cut03} calculated the ``rigidity parameter'' $b$ for a realistic NS structure, with a solid crust afloat on a liquid 
	core. They solved the strain field that develops as the NS spins down and found that $b\sim 10^{-7}$, two orders of magnitude 
	below the result found by  \citet{baym71} for a simplified model. On the other hand, \citet{hor09} recently found through  N-body simulations  that the 
	Coulomb lattice of the  NS crust can support a maximum  strain angle $\theta_{c}\sim 10^{-1}$, three orders of magnitude 
	higher than the value  estimated by \citet{smo70}. Thus, for an NS of $\sim 1.4 M_{\odot}$, the time-averaged crust-cracking luminosity is 
	$L_{cc}\sim  10^{26} \dot P_{-20}/P^3_{5\rm{ms}}~\mathrm{erg~s}^{-1}$, where $\dot P_{-20}$ is the period derivative measured 
	in units of $10^{-20}$ and $P_{5\mathrm{ms}}$ is the period in  units of 5 milliseconds. For a representative classical pulsar
	(PSR B0950+08),	$P\sim 250$~ms and $\dot P\sim 10^{-16}~\mathrm{s~s}^{-1}$, so 
	$L_{cc}\sim 10^{25}~\mathrm{erg~s}^{-1}$, while for the MSP J0437-4715, $P= 5.76$~ms and 
	$\dot P=5.73\times 10^{-20}~\mathrm{s~s}^{-1}$, so  $L_{cc}\sim 4 \times 10^{25}~\mathrm{erg~s}^{-1}$. Comparing these results  
	with the thermal emission from a pulsar with $T_s\sim10^5$~K, $L\sim 10^{29}~\mathrm{erg~s}^{-1}$ (i.e., potentially 
	detectable thermal emission from an NS in the solar neighborhood by the Hubble Space Telescope), we conclude that the crust-cracking mechanism does not  produce detectable heating.

	Additionally, the high critical strain angle obtained by \citet{hor09} requires  the star to have an 
	initial deformation $\epsilon_0> \theta_c\sim 10^{-1} $ in order to cause any cracking of the crust. However, for plausible initial rotation 
	periods of  classical pulsars ($P_0>15$~ms), their initial deformation is only $\epsilon_0 \sim P_K^2/P_0^2 <10^{-3}$, where 
	$P_K\sim 0.5$~ms is the Keplerian period of the star. Hence, the crust-cracking mechanism is never activated in classical pulsars 
	and probably operates only in MSPs with $P_0<2$~ms. If the latter were the case, and now considering that all stresses in the
	crust are suddenly released, the internal thermal energy of the star is increased by $\sim B\theta_c^2\sim 10^{44}$~erg, 
	corresponding to an internal temperature of $10^7$~K and a surface temperature of $\sim 5\times 10^5$~K \citep{pot97}, which
	is dissipated within less than $10^7$~yr. Hence, in MSPs of $10^{8-9}$~yr, the increase in temperature due to catastrophic
	cracking is given by a narrow peak in the thermal evolution, which is unlikely to be detected because of the short timescale 
	involved.

\subsection{Vortex creep}

	The relatively low temperatures in the interior of NSs induce the formation of  neutron Cooper pairs. These form a condensate 
	with a macroscopic wave function. A consequence of this is that the vorticity in the superfluid must be concentrated in discrete
	vortex lines, whose microscopic distribution allows the superfluid to approximate a macroscopic rigid rotation.

	As the star spins down, the vortex lines must move outward. As they move through the inner crust, they are pinned to the 
	nuclear lattice until a critical velocity difference between the superfluid and the crust is reached. In this process, the 
	pinning and unpinning of the vortex lines with respect to the nuclei of the crystal lattice release energy that heats the star. 
	The energy-dissipation rate is given by \citep{alp84}

	\begin{equation}
	L_{vc}=J|\dot\Omega|, 
	\end{equation}
	where $J\simeq \bar \omega I_p$ , with $I_p$  the moment of inertia of the pinning layer and 
	$\bar\omega_{cr}=(\Omega_s - \Omega_c)_{cr}$ an average over the pinning zone of the critical lag between the angular 
	velocity  $\Omega_c$ of the crust and the superfluid rotation rate $\Omega_s$.

	\citet{don04} calculated the vortex-nucleus interaction  in the inner crust 
	of NSs with a semi-classical model. The density-dependent neutron pairing gaps used in the calculations are obtained from the Argonne 
	potential and Gogny effective interaction. Table \ref{tab:uno} shows the pinning energy calculated with this model 
	for five zones of the inner crust. Similarly, \citet{avo08} calculated the vortex-nucleus interaction in the inner crust based 
	on a Hartree-Fock-Bogoliubov  quantum mean field theory. Table \ref{tab:dos} shows the pinning energy for this approach.
        An important result of this, contrary to the  prediction of the  previous model, is that pinning of vortices on nuclei is 
	favored at low density in the inner crust. We used both results to calculate the excess angular momentum 
	$J$, which determines the luminosity for the vortex creep mechanism. In addition, we used Eq. (58) of 
	\citet{alp84} in the limit $E_{NP} \gg kT$,
        \begin{equation}
	\label{eq:jota}
        J=\frac{8\pi}{3} \int_P \frac{E_{NP} r^3}{\kappa \xi R_{WS}} dr
        \end{equation}
        where $E_{NP}$ is the pinning energy of a vortex on a nucleus, $r$ is the radial coordinate, 
	$\kappa$ is the quantum of circulation of each vortex,
	$\xi$ is the vortex coherence length, and $R_{WS}$ is the radius of the Wigner-Seitz cell. In order to calculate 
	this integral, we generate  NS structure models for specific masses and linearly interpolate the values of Table  \ref{tab:uno} 
	and \ref{tab:dos} according to these models. For simplicity, we do not take relativistic effects into account because the 
	corrections involved are minor, i.e. $(1-r_g/R)^{1/2}\sim 0.8$, with $r_g$ the Schwarzschild radius. In this way, for an NS 
	of $M\sim1.4M_\odot$ and a typical range of equations of state we find that the excess of angular momentum is 
	$J\sim (10^{43}-10^{45})$~erg~s. Thus, the vortex-creep luminosity 
	$L_{vc}\simeq (10^{29}-10^{31})|\dot\Omega_{-14}|~\mathrm{erg~s}^{-1}$, where $\dot\Omega_{-14}$ is the angular velocity 
	derivative in units of $10^{-14}~\mathrm{s}^{-2}$. This is similar to the luminosity inferred from the observation of the
	MSP J0437-4715.

        \begin{table}
        \centering
        \caption{Vortex-nuclei interaction parameters in the semi-classical model \citep{don04}, with the Argonne potential (a)  and Gogny potential (b).}
        \begin{tabular}{c c c c c c}
        \noalign{\smallskip}
        \hline
        \noalign{\smallskip}
	Zone & $\rho~[\mathrm{g~cm}^{-3}]$ & $R_{WS}~[\mathrm{fm}]$  & $\xi~[\mathrm{fm}]$ & $E^{(a)}_{NP}~[\mathrm{MeV}]$ & 
	$E^{(b)}_{NP}~[\mathrm{MeV}]$\\
        \noalign{\smallskip}
        \hline
        \noalign{\smallskip}
        1  & $1.5\times10^{12}$  & $44.0$ & $6.54$  &  $-$      & $-$\\
        2 & $9.6\times10^{12}$  & $35.5$ & $7.25$  &  $-$       & $-$\\
        3 & $3.4\times10^{13}$  & $27.0$ & $8.54$  &  $5.2$     & $-$\\
        4 & $7.8\times10^{13}$  & $19.4$ & $11.71$ &  $5.1$     & $7.5$\\
        5 & $1.3\times10^{14}$  & $13.8$ & $8.62$  &  $0.4$     & $5.9$\\
        \hline
        \end{tabular}
	\tablefoot{$\rho$ is the density of each zone, $R_{WS}$ is the radius of the Wigner-Seitz cell, $\xi$ is the
	vortex coherence length, and $E_{NP}$ is the vortex-nuclei  pinning energy.}
	\label{tab:uno}
        \end{table}
        \begin{table}
        \centering
        \caption{Vortex-nuclei interaction parameters in the quantum approach \citep{avo08}, with the SLy4 potential (a) and Skm* potential (b).}
        \begin{tabular}{c c c c c c}
        \noalign{\smallskip}
        \hline
        \noalign{\smallskip}
        Zone & $\rho~[\mathrm{g~cm}^{-3}]$  & $R_{WS}~[\mathrm{fm}]$  & $\xi~[\mathrm{fm}]$ & $E^{(a)}_{NP}~[\mathrm{MeV}]$ & 
	$E^{(b)}_{NP}~[\mathrm{MeV}]$\\
        \noalign{\smallskip}
        \hline
        \noalign{\smallskip}
        1  & $1.9\times10^{12}$  & $42.9$ & $6.63$  &  $1.08$      & $1.51$\\
        2 & $3.3\times10^{12}$  & $40.3$ & $6.85$  &  $1.20$       & $3.85$\\
        3 & $6.6\times10^{12}$  & $37.2$ & $7.11$  &  $-$     & $1.63$\\
        4 & $1.3\times10^{13}$  & $33.2$ & $7.60$ &  $-$     & $-$\\
        \hline
        \end{tabular}
	\tablefoot{ Variables are as explained for Table \ref{tab:uno}. The values of $\xi$ are interpolated from Table 8
		of \citet{don04}.}
	\label{tab:dos}
        \end{table}
	
\subsection{Rotochemical heating}
	
	In chemical equilibrium, in an NS composed of neutrons ($n$), protons ($p$), and leptons ($l$: electrons and muons), the 
	chemical potentials satisfy	$\eta_{npl}\equiv \mu_{n}-\mu_{p}-\mu_{l}=0$. However, if the rotation of the star 
	is slowing down, the centrifugal force is reduced, the central density of the star increases, and the chemical potentials are 
	imbalanced, $\eta_{npl} \neq0$. As the equilibrium composition is altered, the NS will relax to the new chemical equilibrium 
	(via beta and inverse beta decays), releasing  energy in the form of neutrinos, which leave the star, and heat, which is
	later radiated as photons. The evolution of the chemical imbalances is of the form  \citep{fer05}
       	\begin{equation}
	\label{eq:eta}
       	\dot \eta_{npl}= -A(\eta_{npl},T)-R_{npl}\Omega\dot\Omega,
       	\end{equation}
 	where the function $A$ quantifies the effect of reactions toward restoring chemical equilibrium, and the scalar $R_{npl}$ 
	quantifies the departure from equilibrium due to  the change in the angular velocity $\Omega$ of the star.
	
	The luminosity generated by this mechanism is $L= \Gamma\eta_{npl}$, where $\Gamma= \Gamma_{n\to pl\bar\nu}-\Gamma_{pl\to n\nu}$.
	Here, $\Gamma_{n\to pl\bar\nu}$ is the rate of reactions (integrated over the core) that transform the neutrons to protons and 
	leptons through direct or modified Urca reactions, and $\Gamma_{pl\to n\nu}$ is the rate for the opposite process. In  this way, the evolution of the internal temperature  with 
	rotochemical heating is given by the solution of the coupled differential Eqs. (\ref{eq:ther_bal}) and (\ref{eq:eta}).

	\citet{reis95} found that if the angular velocity  $\Omega$ varies slowly over the time required to cool the star and achieve 
	chemical equilibrium, the star reaches a quasi-steady state, where heating and cooling are balanced. \citet{fer05} 
	calculated the simultaneous solution of  $\dot T = \dot \eta_{npl} = 0$	for a typical range of equations of state  and 
	found that in an NS with a non-superfluid core and with modified Urca reactions, the photon luminosity in the quasi-steady state
	depends only on the period and its derivative,
	\begin{equation}
	L_{\gamma}^{st} \simeq (10^{30}-10^{31})\left(\frac{\dot{P}_{-20}}{P_{\rm{ms}}^3}\right)^{8/7}~\mathrm{erg~s}^{-1},
	\end{equation}
	which is close to matching the observation of MSP J0437-4715.
	Here, $\dot P_{-20}$ is the period derivative measured in units of $10^{-20}$ and $P_{\rm{ms}}$ is the period in milliseconds. 
	The characteristic timescale  to reach this quasi-steady state is 
	$\tau_{eq} \simeq 2 \times 10^7 (P_{\rm{ms}}^{3}/\dot P_{-20})^{6/7}$~yr.
	On the other hand, \citet{petro09} considered the effects of nucleon superfluidity on rotochemical heating. They found that the
	chemical imbalances grow up to a value close to the energy gaps, which is higher than in  the nonsuperfluid case. 
	Therefore, the surface temperatures predicted with Cooper pairing are higher. For simplicity, we here only consider 
	the non-superfluid case. However, we must not lose sight of the latter result.			

\section{Effects}\label{sec:eff}
		
	In order to solve the thermal balance equation, Eq. (\ref{eq:ther_bal}), and generate the evolutionary curves, we use the code  of 
	\citet{fer05}. This considers realistic equations state (EOSs),  with  a conventional NS divided into two regions: an isothermal 
	interior and a thin envelope. In the interior, it considers a  core composed of neutrons, protons, electrons, and muons, but
	ignores the potential Cooper pairing effect. The neutrino emissivity is generated by modified Urca reactions and direct Urca 
	reactions, neglecting any other neutrino emission processes. In order to model the envelope, it uses the relation between 
	internal and surface temperature from the fully accreted envelope model of \citet{pot97}. The rotational evolution  is assumed 
	to be due to magnetic dipole radiation, without magnetic field decay.

\subsection{Vortex creep}\label{sec:vc}% effects
	
	\begin{figure}[t]%1%
	\centering
        \includegraphics[width=9cm,height=6cm]{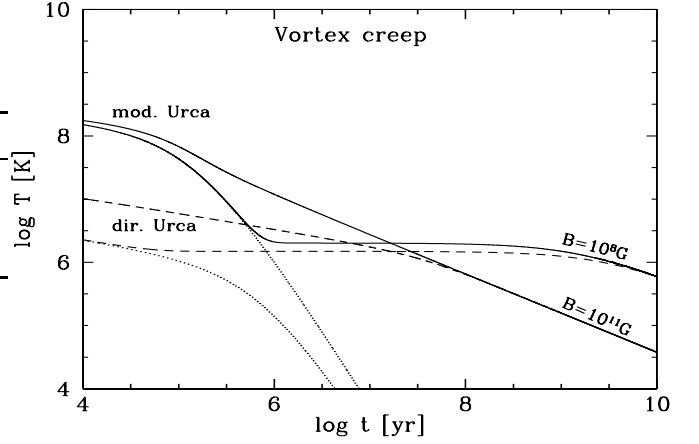}
	\caption{Thermal evolution with the vortex creep mechanism. All curves show the interior temperature as a function of time, for
	stars with mass $M=1.4M_{\odot}$, initial temperature $T=10^{11}$~K, and initial period $P_0=1$~ms. The magnetic fields 
	$B=10^8$~G and $B=10^{11}$~G correspond to MSPs and classical pulsars, respectively. The excess angular momentum used is  
	$J=10^{43}~\mathrm{erg~s}$. The solid lines show the evolution using the  A18+ $\delta \upsilon$ + UIX* EOS, 
	with only modified Urca reactions, and the dashed lines show the evolution using BPAL 22 EOS, with direct Urca reactions.
	The dotted lines show the evolution with passive cooling for direct and modified Urca reactions.} 
        \label{fig:cp_vc}
        \end{figure}

	Figure \ref{fig:cp_vc} shows the resulting thermal evolution for the vortex creep mechanism for representative parameters of both
	classical pulsars and MSPs, with one EOS allowing only for modified Urca reactions (A18+$\delta v$+UIX* EOS, \citealt{akm98}) and  
	another one allowing for direct Urca reactions (BPAL 22, \citealt{prak88}). As the residual spin-down at old ages is greater in 
	pulsars with weaker magnetic fields, the predicted temperature is higher in the MSPs than 
	in the classical pulsars. 
	At very late times ($t>10^8$~yr), the cooling is dominated by photon emission, which balances the heat generation by vortex 
	friction. Therefore, the evolution of temperature becomes independent 	of the previous thermal history and the type of Urca 
	reaction, and only depends on the current value of $\dot\Omega$, with the photon luminosity 
	$L=4\pi R^2\sigma T_s^4 = J |\dot\Omega|$.  The rotational	evolution by magnetic dipole radiation yields 
	$\Omega\propto t^{-1/2}$ in the limit $\Omega\ll\Omega_0$. Thus, the asymptotic 
	decrease of the surface temperature with time in classical pulsars and MSPs is $T_s\propto t^{-3/8}$.

\subsection{Rotochemical heating}\label{sec:rq}%effects				
	
     	\begin{figure}[t]%2%
        \centering
        \includegraphics[width=9cm,height=6cm]{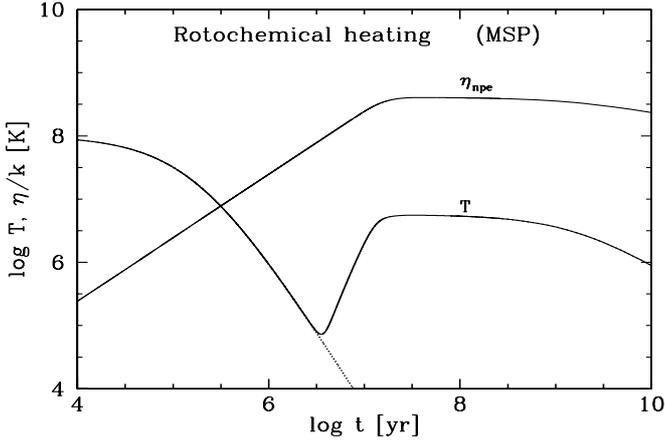}
        \caption{Thermal evolution with rotochemical heating. All curves correspond to stars with mass $M=1.4M_{\odot}$, 
	A18 + $\delta \upsilon$ + UIX* EOS (which allows only for modified Urca reactions), initial temperature $T=10^{8}$~K, and  magnetic field $B=10^8$~G. The solid lines 
	show the evolution of the internal temperature $T$ and the chemical imbalance $\eta_{npe}$ , and the dotted line shows 
	the  passive cooling.}
        \label{fig:rq_msp}
        \end{figure}

        \begin{figure}%3%
        \centering
        \includegraphics[width=9cm,height=6cm]{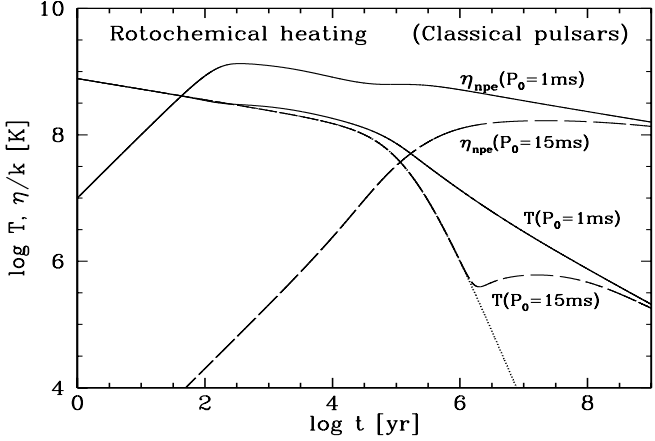}
        \caption{Thermal evolution with rotochemical heating. All curves correspond to stars with mass $M=1.4M_{\odot}$,
	A18 + $\delta \upsilon$ + UIX* EOS, initial temperature $T=10^{11}$~K, and magnetic field $B=10^{11}$~G.
	The solid and long-dashed lines show the evolution of the temperature and the chemical imbalance with  initial
	periods $P_0=1$~ms and $P_0=15$~ms, respectively. The dotted line shows the passive cooling.}
        \label{fig:rq_cp}
        \end{figure}

	The thermal evolution for MSPs with rotochemical heating and modified Urca reactions is shown in Fig.  \ref{fig:rq_msp}. For 
	this mechanism, the thermal evolution is coupled with the evolution of the chemical imbalances. Because the magnetic fields in MSPs 
	are relatively weak, $B\sim10^8$~G, the chemical imbalances induced by the spin-down grow slowly, causing chemical reactions at 
	high ages, $t\ga10^7$~yr. An important prediction of this mechanism is that in the MSP regime, the star arrives at a 
	quasi-steady state, where heating and cooling balanced,  so the thermal evolution is independent of initial conditions and 
	only depends on the current value of the product $\Omega\dot\Omega$ \citep{fer05}. When direct Urca reactions are present, the
	evolution is qualitatively similar, but the
	temperatures  are strongly reduced, as shown  in Sect. \ref{sec:obs}.  

	The evolution of classical pulsars is shown in Fig. \ref{fig:rq_cp} \citep[see also][]{reis07}. As in the previous figure, only 
	modified Urca processes are active. Owing to the relatively strong magnetic fields present in this regime, $B\ga10^{11}$~G, the 
	rotational energy of the star is quickly consumed in the beginning of the thermal evolution  by the  magnetic dipole radiation. 
	Because of this, high chemical imbalances are built up, proportional to the amount of rotational energy lost. Thus, for faster 
	initial rotation, the chemical imbalance at later times will be higher.
	At ages above $\sim 10^8$~yr, the right-hand side of Eq. (\ref{eq:eta}) is  dominated by the reactions restoring the chemical
	equilibrium. When only  modified Urca reactions are present and $\eta \gg kT$, the chemical imbalance evolves according to 
	$\dot\eta\propto - \Gamma \propto-\eta^7$ \citep[see][]{fer05}, yielding $\eta \propto t^{-1/6}$. Because in the last epoch of the 
	thermal evolution the cooling of the star is caused mainly by photon emission, the luminosity is 
	$L\propto T_s^4\propto\Gamma\eta \propto \eta^8$. Thus, 
	the surface temperature decreases with age as $T_s\propto t^{-1/3}$, only slightly more slowly than with vortex creep.
\section{Comparison with observations}\label{sec:obs}

	In order to  identify  the thermal emission at $T\sim 10^5$~K from the whole NS surface, it is necessary to obtain 
	observations in the optical  and specially in the ultraviolet range \citep{kar04,zav04}. Thus, to confront the theoretical 
	cooling models of old NSs with observations, we select the best and most restrictive observations in these bands, of seven very 
	old pulsars: three MSPs and four classical pulsars. Among the MSPs, only  J0437-4715 has detected thermal emission 
	\citep{kar04}, and J2124-3358 \citep{mig04} and J0030+0451 \citep{kop03}  have good upper limits on their temperatures. The four selected 
	classical pulsars, B1929+10 \citep{beck06}, B0950+08 \citep{zav04,zhar04}, B1133+16 \citep{zhar08},  
	and J0108-1431 \citep{mig03} only  have upper limits on the thermal emission from the whole NS surface.

\begin{figure}[t]%4%
	\includegraphics[width=9cm,height=5.5cm]{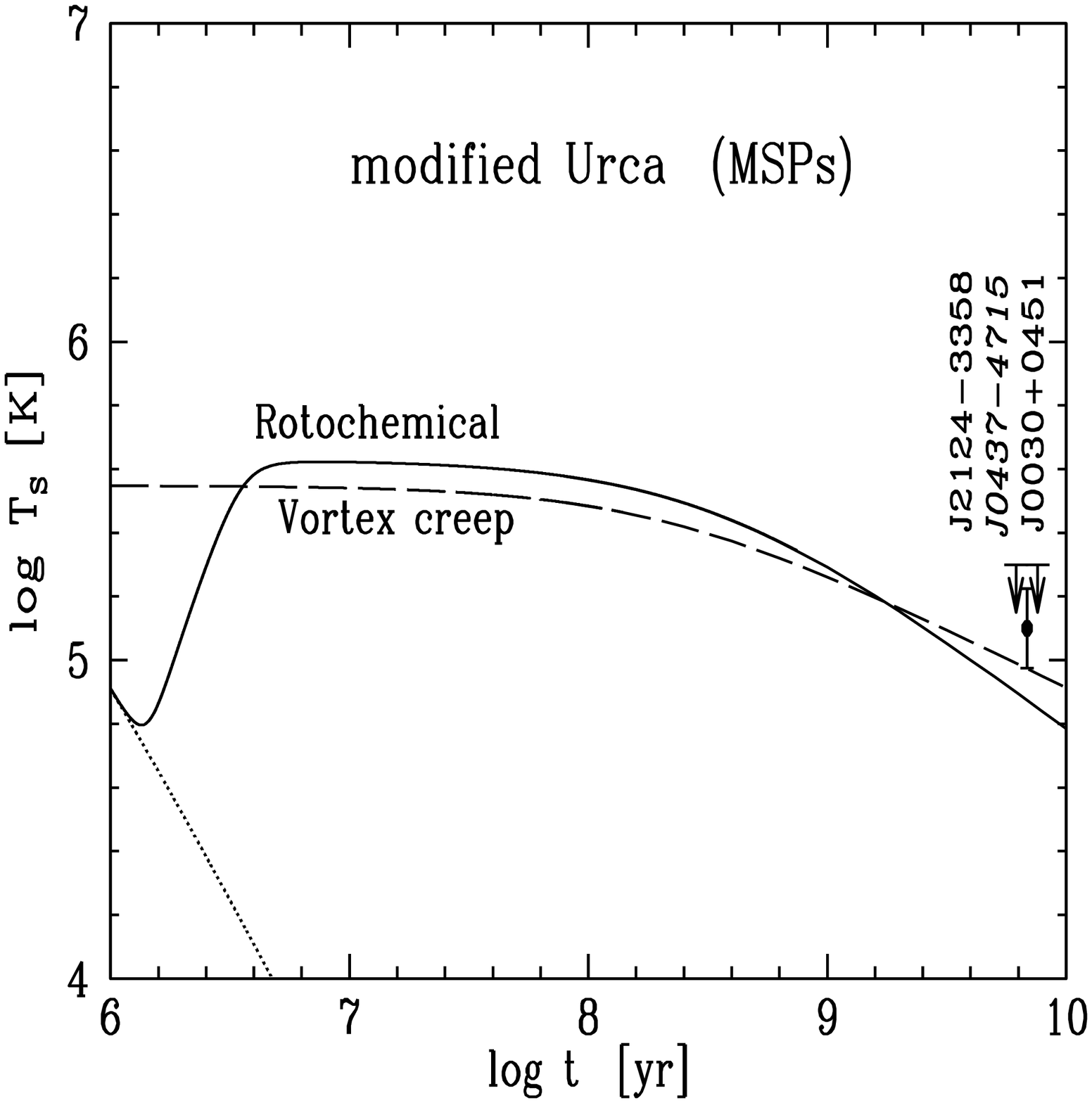}
        \includegraphics[width=9cm,height=5.5cm]{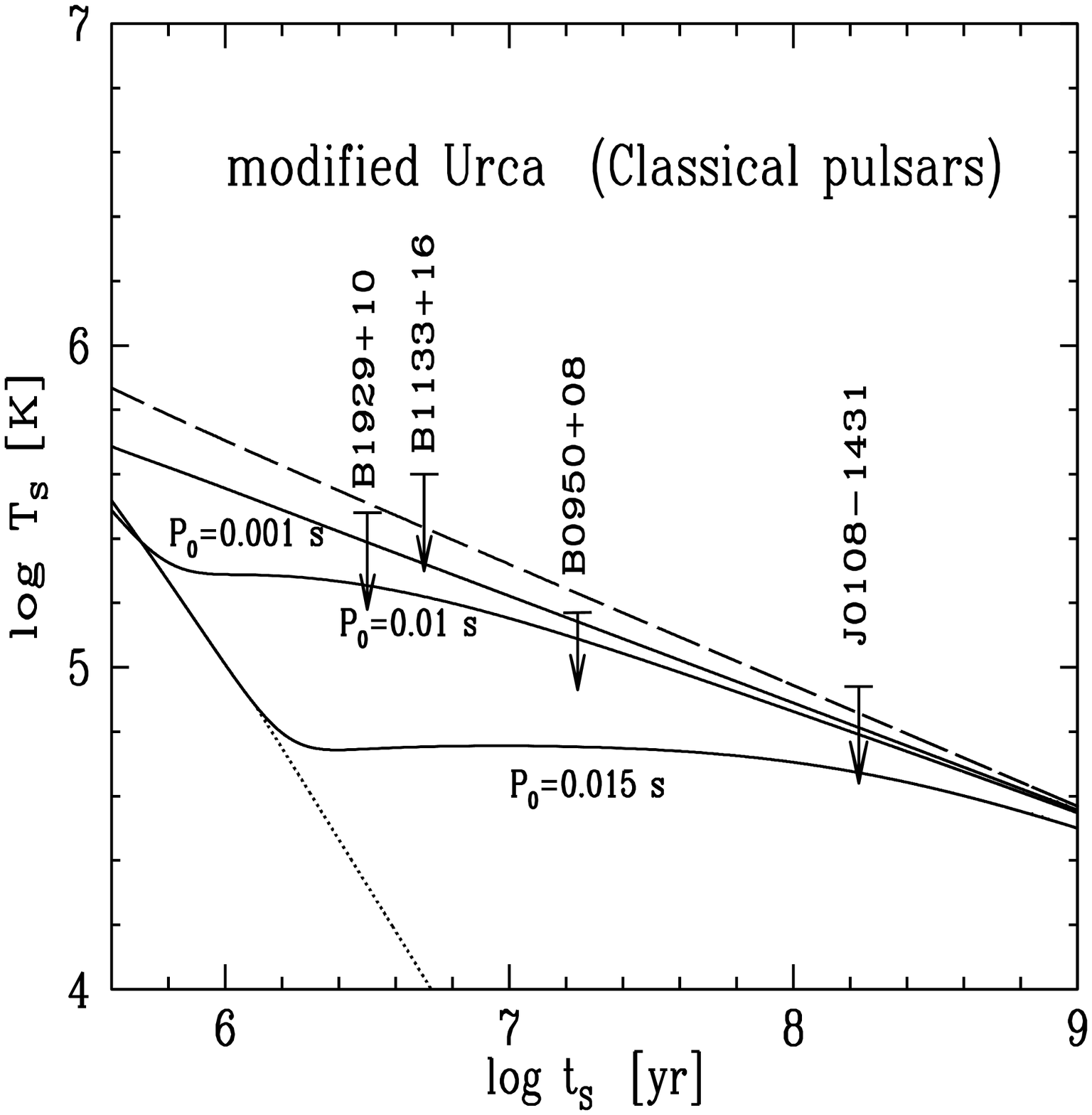}
	\caption{ Evolution of the surface temperature for a neutron star with vortex creep (long-dashed lines),
        rotochemical heating (solid lines), and passive cooling (dotted lines). All curves correspond to stars
	with the A18 + $\delta \upsilon$ + UIX* EOS and modified Urca reactions. The error bar shows the temperature measured 
	for the MSP J0437-4715 and the arrows show the upper limits on the thermal emission for specific pulsars.
        {\bf Top panel}: The curves correspond to MSPs with mass $M=1.76M_{\odot}$, magnetic field  $B=2.8\times10^{8}$~G,  
	and initial temperature $T=10^8$~K. 
	{\bf Bottom panel}: The curves correspond to classical pulsars with mass $M=1.4M_{\odot}$, magnetic field 
	$B=2.5\times10^{11}$~G, and initial temperature $T=10^{11}$~K. The abscissa corresponds to the spin-down time  
	($t_s=\Omega/2|\dot\Omega|$). The initial periods for rotochemical heating are labeled on each curve.}
        \label{fig:teo_obs1}
        \end{figure}

 	The upper panel of Fig. \ref{fig:teo_obs1} shows the thermal evolution of MSPs including vortex creep and rotochemical heating 
	with modified Urca reactions. For the vortex creep mechanism the parameter $J$ is highly uncertain, so we adjust it to the lowest value
	that is consistent with the thermal emission from MSP J0437-4715.
	This constrains it to $J\geq5.5\times10^{43}$~erg~s if vortex creep is postulated as the main heating 
	mechanism for this pulsar. For rotochemical heating, the surface temperature of this MSP can be quite precisely predicted, 
	considering its mass $M=1.76M_{\odot}$ \citep{ver08} and an interior model given by the  A18 + $\delta \upsilon$ + UIX* EOS. 
	This prediction is $1.7 \sigma$ below the observation of \citet{kar04}.

        \begin{figure}[t]%5%
        \includegraphics[width=9cm,height=5.5cm]{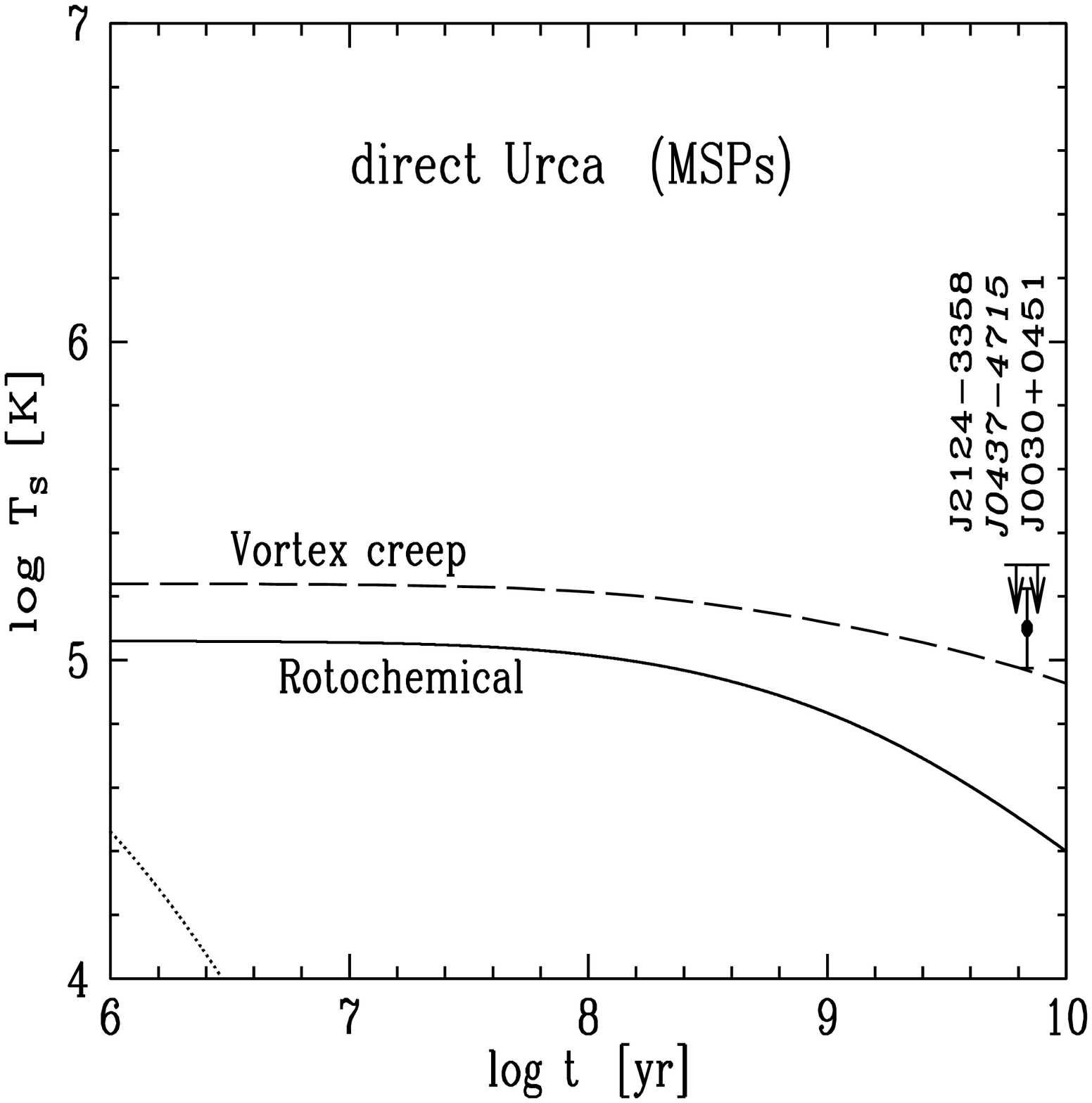}
	\includegraphics[width=9cm,height=5.5cm]{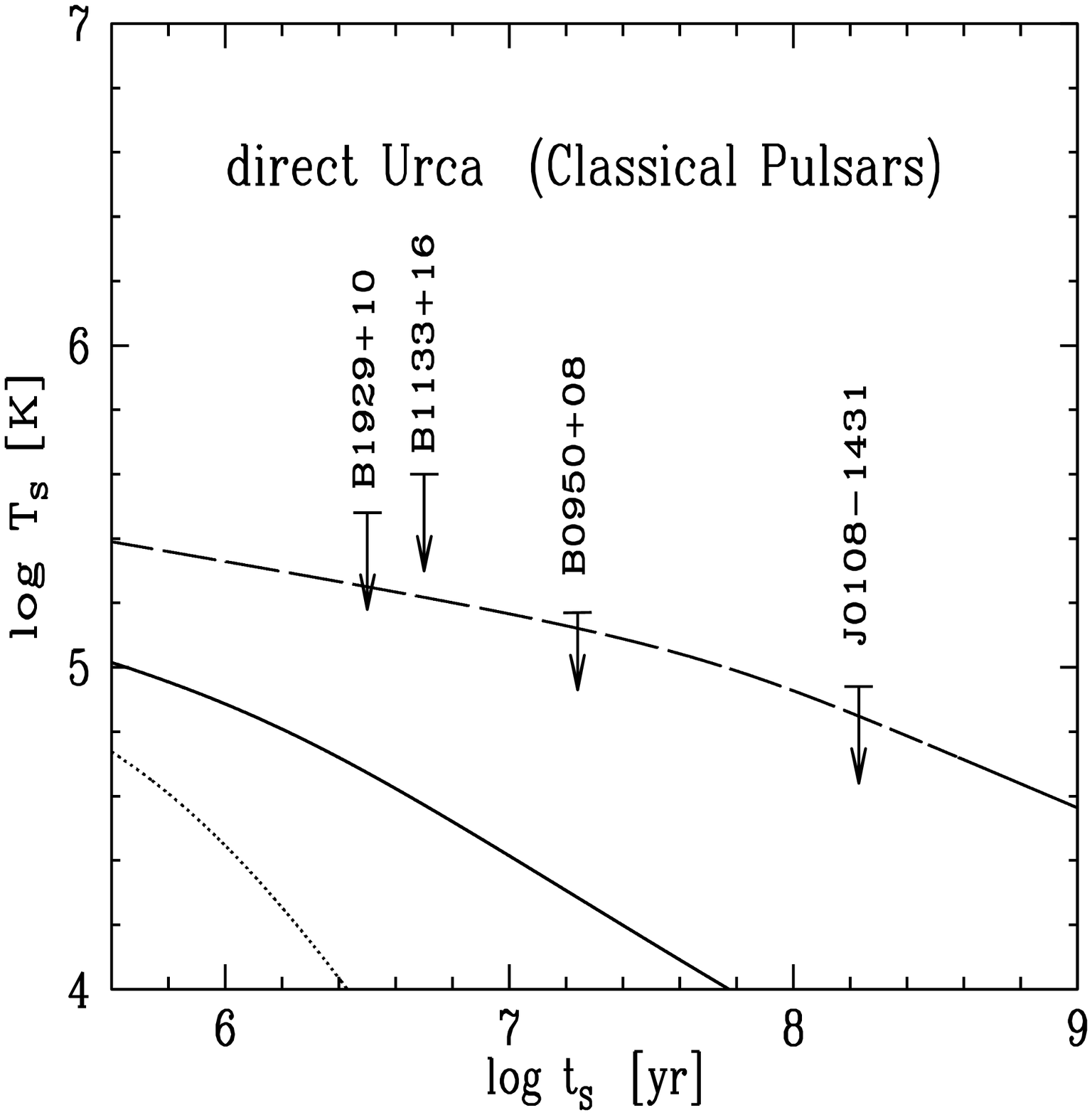}
	\caption{ Evolution of the surface temperature for a neutron star with vortex creep (long-dashed lines),
        rotochemical heating (solid lines), and passive cooling (dotted lines). All curves correspond to stars
	with the BPAL 32 EOS and direct Urca reactions. The error bar shows the temperature measured 
	for the MSP J0437-4715 and the arrows show the upper limits on the thermal emission for specific pulsars.
         {\bf Top panel}: The curves correspond to MSPs with mass $M=1.76M_{\odot}$, magnetic field  $B=2.8\times10^{8}$~G,  
	and initial temperature $T=10^8$~K. 
        {\bf Bottom panel}: The evolutionary curves correspond to stars with mass $M=1.4M_{\odot}$, magnetic field 
	$B=2.5\times10^{11}$~G, and initial temperature $T=10^{11}$~K. The abscissa corresponds to the spin-down time 
	($t_s=\Omega/2|\dot\Omega|$). The initial period for rotochemical heating is $P_0=1$~ms.}
        \label{fig:teo_obs2}
        \end{figure}

	Similarly, the bottom panel shows the thermal evolution of the classical pulsars. For the vortex creep mechanism, we assumed 
	that the parameter $J$ is universal. Thus, to solve the thermal evolution in this regime, we used the previous value of 
	$J$ imposed by the MSP J0437-4715, resulting in a ``phenomenological'' temperature prediction slightly above the constraints for the
	pulsars B1929+10 and B0950+08. On the other hand, as we saw in Sect.  \ref{sec:rq}, the evolution with rotochemical heating 
	depends on the initial period  of rotation, $P_0$, for NSs with relatively high magnetic fields. Because of this, rotochemical 
	heating can easily accommodate substantially lower temperatures if more restrictive observations of classical pulsars 
	are available. 	The evolution at late times ($t>10^8$~yr) is remarkably similar for both mechanisms. 
	We showed that the relation between surface temperature and age is   $T_s \propto t^{-3/8}$ for vortex creep 
	(Sect. \ref{sec:vc}), very slightly steeper than the relation $T_s \propto t^{-1/3}$ for rotochemical heating  with modified 
	Urca reactions (Sect. \ref{sec:rq}) in the latter stage of the thermal evolution. 
		
	Figure \ref{fig:teo_obs2} shows the thermal evolution considering the fast direct Urca processes. For the vortex creep mechanism, 
	we used the same excess of angular momentum $J$ as in the previous evolutionary curves, fixed to be consistent with the 
	observed temperature of MSP J0437-4715. Because for ages $\ga10^8$yr this 
	mechanism does not depend on the Urca process type, the temperatures of the MSPs (upper panel) are the same as with modified 
	Urca reactions. However, in classical pulsars (bottom panel) the thermal evolution is sensitive  to the direct Urca processes 
	and generates a predicted  temperature  slightly lower than the limit for PSR B0950+08. In this way, a
	more sensitive observation of this pulsar could discard vortex creep as the main source of the thermal emission detected from 
	the MSP J0437-4715. On the contrary, the temperatures generated by rotochemical heating, both for classical pulsars and MSPs,
	are strongly reduced if direct Urca processes are considered. The temperature measured in J0437-4715 requires that the neutrino emission is 		produced only by modified Urca reactions \citep[unless substantial	Cooper pairing gaps are present;][]{petro09} if rotochemical 
	heating is  the main source of heat.

	Finally, we compare the excess angular momentum, $J$, that we estimated from the pinning energies of Tables \ref{tab:uno} and \ref{tab:dos}
	with the observations. In order to do this, we numerically integrated Eq. \ref{eq:jota}  over the inner crust of an NS with 
	canonical mass $M=1.4M_{\odot}$ and the A18+$\delta v$+UIX* EOS. In this way, with the pinning energy estimated by \citet{don04}, 
	we obtain $J=3.8\times10^{44}~\mathrm{erg~s}$ for the Argonne interaction and $J=5.9\times 10^{44}~\mathrm{erg~s}$ for the 
	Gogny interaction. Similarly, but with the pinning energy estimated by \citet{avo08}, $J=1.2\times10^{43}~\mathrm{erg~s}$ with 
	the SLy4 interaction, and $J=6.8\times 10^{43}$~erg s with the Skm* interaction. Comparing these results with the value 
	$J = 5.5\times 10^{43}$~erg~s, which is  marginally compatible with both the observed temperatures of MSP J0437-4715 and the upper limits
	for the old classical pulsars, we find that the semi-classical model of \citet{don04} overestimates the value of the excess 
	angular mometum $J$, while the quantum approach of \citet{avo08}, with the SLy4 interaction, roughly agrees with the 
	value inferred from observations.

\section{Conclusions}\label{sec:conc}

	We studied five heating mechanisms that can be operating in old neutron stars: magnetic field decay, dark matter
	accretion, crust cracking, vortex creep, and rotochemical heating, and compared them with pulsar observations.

	We found that magnetic field decay, dark matter accretion, and crust cracking cannot produce detectable heating in
	old pulsars. 
	Owing to the high yield strain angle \citep{hor09}, the crust cracking mechanism does not operate in classical pulsars, and 
	probably only operates in MSPs. The vortex creep and rotochemical heating can be important both for classical pulsars 
	and MSPs. 

	In the evolutionary curves with vortex creep, and with the excess angular  momentum, $J$, set to the lowest temperature 
	compatible with the thermal emission detected from MSP J0437-4715, the predicted  temperature turns out to be very near 
	the observational upper limits for the classical pulsars B1929+10 and B1133+16. Likewise, rotochemical heating
	with modified Urca reactions and no Cooper pairing is only 1.7 $\sigma$ below the temperature measured in the  
	MSP J0437-4715. However, the temperature prediction of rotochemical heating can be raised if a superfluid core is considered 
	in the model. 

	The prediction of pinning energies in the inner crust of NSs via the semi-classical model for the vortex-nuclei interaction 
	\citep{don04} overestimates the temperatures  of B0950+08 and B1929+10. The recent estimations of pinning energies via a quantum 
	approach \citep{avo08} are consistent with the observations of classical pulsars and MSPs. Finally, more stringent constraints 
	on the temperature of some classical pulsars such as B0950+08 could rule out the vortex creep mechanism as the main source of 
	the thermal emission detected in the MSP J0437-4715.

	\begin{acknowledgements}
	We thank  Sebasti\'an Reyes, Ricardo Ram\'irez, and Miguel Kiwi  for discussions and comments that benefited this paper, 
	and Rodrigo Fern\'andez for letting us use his rotochemical heating code.
	This work was supported by Proyecto Regular FONDECYT 1060644, the FONDAP Center of Astrophysics (15010003),
	ALMA-CONICYT project 31070001, Proyecto Basal PFB-06/2007, Gemini-CONICYT project
	32080004, and Proyecto L\'imite VRAID $\rm{N}^o$ 15/2010.
	\end{acknowledgements}
	%%%%%%%%%%%%%%%%%%%%%%%%%%%%%%%%%%%%%%%%%%%%%%%%%%%%%%%%%%%%%%%%%%%%%%%%%%%%%%%%%%%%%%%%%%%%%
%	\bibliographystyle{aa}
	
	\end{document}